\documentclass[12pt]{article}
\usepackage{epsfig}
\usepackage{cite}
\usepackage[hyperindex,breaklinks]{hyperref} 
\usepackage{amsmath,xfrac}
\begin{document}

\begin{titlepage}

\vspace{1.5cm}

\begin{center}
\ 
\\
{\bf\large Looking for magnetic monopoles at  LHC with diphoton events}
\\
\date{ }
\vskip 0.70cm

Luis N.\ Epele$^{a}$, Huner Fanchiotti$^{a}$, Carlos A.\ Garc\'{\i}a
Canal$^{a}$, Vasiliki A.\ Mitsou$^{b}$
\\ and Vicente Vento$^{b,c}$

\vskip 0.30cm

{$^{(a)}$ \it Laboratorio de F\'{\i}sica Te\'{o}rica, Departamento de
F\'{\i}sica, IFLP, CONICET,\begin{scriptsize}\begin{footnotesize}\end{footnotesize}\end{scriptsize} \\ Facultad de Ciencias Exactas, Universidad
Nacional de La Plata
\\C.C. 67, 1900 La Plata, Argentina.}\\({\small
E-mail: epele@fisica.unlp.edu.ar, huner@fisica.unlp.edu.ar, garcia@fisica.unlp.edu.ar})

\vskip 0.3cm
{$^{(b)}$ \it Instituto de F\'{\i}sica Corpuscular\\
 Universidad de Valencia and CSIC\\
Apartado de Correos 22085, E-46071 Valencia, Spain.}\\
({\small E-mail:
vasiliki.mitsou@ific.uv.es})\\
\vskip 0.3cm 
{$^{(c)}$ \it Departamento de F\'{\i}sica Te\'orica \\
Universidad de Valencia \\
E-46100 Burjassot (Valencia), Spain.} \\ ({\small E-mail:
vicente.vento@uv.es}) 
\end{center}

\vskip 1cm 
%\centerline{\bf Abstract}
\begin{abstract}
Magnetic monopoles have been a subject of interest since Dirac established the
relation between the existence of monopoles and charge quantization.
The intense experimental search carried thus far has not met with success. 
The Large Hadron Collider is reaching energies never achieved before allowing 
the search for exotic particles in the TeV mass range. In a continuing effort
to discover these rare particles we propose here other ways to detect them.
We  study the observability  of monopoles and monopolium,  a monopole-antimonopole 
bound state, at the Large Hadron Collider  in the $\gamma \, \gamma$ channel for 
monopole masses in the range 500--1000 GeV.  We conclude that  LHC is an ideal machine to
 discover monopoles with masses  below 1 TeV at present running energies and with 5~fb$^{-1}$
 of integrated luminosity.
\end{abstract}

 \vspace{1cm}

\noindent Pacs: 14.80.Hv, 95.30.Cq, 98.70.-f, 98.80.-k


\end{titlepage}

\section{Introduction}

The theoretical justification for the existence of classical magnetic poles, hereafter called
monopoles, is that they add symmetry to Maxwell's equations and explain charge 
quantization \cite{Dirac:1931kp,Jackson:1982ce}. Dirac showed that the mere existence 
of a monopole in the universe could
offer an explanation of the discrete nature of the electric charge. His analysis leads to the
 Dirac Quantization Condition (DQC),

\begin{equation} e \, g = \frac{N}{2} \;, \mbox{  N = 1,2,...}\;, 
\label{dqc}\end{equation}

\noindent where $e$ is the electron charge, $g$ the monopole
magnetic charge and we use natural units $\hbar = c =1$.
In Dirac's formulation, monopoles are assumed to exist as point-like particles and quantum mechanical
consistency conditions lead to Eq.~(\ref{dqc}), establishing the value of their magnetic charge.
Their mass, $m$, is a parameter of the theory.

Monopoles and their experimental detection
have been a subject of much study since many believe in Dirac's statement \cite{Dirac:1931kp},

\begin{center}
{\it ``...one would be surprised if Nature had made no use of it [the monopole]."}
\end{center}
\vskip 0.2cm
All experimental searches for magnetic monopoles up to now have met with failure
\cite{Craigie:1986ws,Martin:1989ms,Abbott:1998mw,Eidelman:2004wy,Mulhearn:2004kw,
Giacomelli:2005xz,Abulencia:2005hb,Milton:2006cp,Yao:2006px, 
Balestra:2011ks,Abulencia:2005kq}. These experiments have led to a lower mass limit in the range of 
$350$ GeV. The lack of experimental confirmation has led many physicists to
abandon the hope in their existence. 

Although monopoles symmetrize Maxwell's equations in form, there is a numerical
asymmetry arising from the DQC, namely that the basic magnetic charge is much larger
than the smallest electric charge. This led Dirac himself in his 1931 paper \cite{Dirac:1931kp} to state,

\begin{center}
\begin{minipage}{6in}
{\it ``... the attractive force between two one-quantum poles of opposite sign is $( 137/2 )^2 \approx 4692\sfrac{1}{4}$ times that between the electron and the proton. This very large force may perhaps
account for why the monopoles have never been separated."}
\end{minipage}
\end{center}
\vskip 0.2cm

Inspired by this old idea of Dirac and  Zeldovich 
\cite{Dirac:1931kp,Dirac:1948um,Zeldovich:1978wj}, namely, that
monopoles are not seen freely because they are confined by their
strong magnetic forces forming a bound state called monopolium
\cite{Hill:1982iq,Dubrovich:2002gp}, we proposed that monopolium,  due to its 
 bound state structure,  might be easier to detect than free monopoles  
\cite{Epele:2007ic,Epele:2008un}.

The Large Hadron Collider (LHC), which entered last year  in operation colliding  $3.5$-TeV protons,
will probe the energy frontier opening possibilities for new physics including the discovery of magnetic monopoles either
directly, a possibility contemplated long time ago
\cite{Ginzburg:1981vm,Ginzburg:1982fk}, and revisited frequently \cite{Abbott:1998mw,Abulencia:2005hb,Kurochkin:2006jr,Ginzburg:1998vb,Ginzburg:1999ej,Kalbfleisch:2000iz,Dougall:2007tt}
or through the discovery of monopolium, as advocated in refs. \cite{Epele:2007ic,Epele:2008un}. 
The direct observation is based on the search for highly ionizing massive particles at the ATLAS~\cite{atlas}
and CMS~\cite{cms} detectors or at the MoEDAL experiment~\cite{moedal}, which is
designed to search precisely for such exotic states. 

If instead of single monopoles (antimonopoles) we deal with monopole-antimonopole pairs, as is the case at LHC, we expect that due to the very strong interaction many of them annihilate into photons inside the detector, either directly or by forming a monopolium bound state which will also disintegrate producing  photons. Therefore we are led to study  the production of monopoles and monopolium at  LHC by the mechanism of photon fusion and its  subsequent decay into two photons \cite{Epele:2011cn,Epele:2011pn}.

In the next section we describe the dynamics of monopoles and review the production of monopole-antimonopole pairs.
Section~\ref{dynamics} discusses  the annihilation of the virtual monopole-antimonopole pair into photons. In section~\ref{monopole} we review the production of monopolium and study its annihilation into photons. In section ~\ref{monopolium} we describe how to incorporate the elementary 
processes into $p-p$ scattering. In section~\ref{results} we present our results in the context of the present running features of LHC and in the last section we draw conclusions of our studies.

\section{Monopole dynamics}\label{dynamics}

The theory of monopole interactions was initially formulated by Dirac \cite{Dirac:1948um} and later on developed in two different approaches
by Schwinger \cite{Schwinger:1966nj} and Zwanziger \cite{Zwanziger:1970hk}.  The formalism of Schwinger can be cast in 
functional form as a field theory for monopole-electron interaction which is dual to 
Quantum Electrodynamics (QED) \cite{Gamberg:1999hq}. The monopoles are considered fermions and behave as 
electrons in QED with a large coupling constant as a result of the DQC. In this formulation  the conventional photon 
field is Dirac string dependent. Due to the large coupling constant and the string 
dependence, perturbative treatments \`a la Feynman are in principle 
not well defined. However, non perturbative high energy treatments, like the eikonal approximation, 
have rendered well defined electron-monopole cross sections \cite{Gamberg:1999hq,Urrutia:1978kq}.
For the case of monopole production at energies higher than their mass the above procedure is not 
applicable, and being the treatment non perturbative, there is no universally accepted prediction 
from field theory.  

A different scheme to understand monopole interactions was proposed  by Ginzburg and Schiller  \cite{Ginzburg:1998vb,Ginzburg:1999ej}. 
Using standard electroweak theory in the one-loop approximation  to lowest non-vanishing order they derived an effective theory with
coupling,

\begin{equation}
g_{\textit{eff}} = \frac{g \omega}{\sqrt{4 \pi}  m},
\end{equation}
where $g$ is the monopole coupling, $\omega$ the photon energy of the vertex and $m$ the monopole mass. Thus, for virtualities smaller than $\omega$ and $\omega << m$, one can derive an effective Heinsenberg-Euler type effective Lagrangian for the box diagram with a small coupling  \cite{Ginzburg:1998vb}.  However, one can always ask the question: what will happen at higher orders? How can one treat the divergencies? The authors argue that in this kinematical region general considerations like gauge invariance,
threshold behavior, etc., together with the perturbative approach allows an effective  theory analogous to standard 
QED  in lowest non trivial order only.

Certainly their procedure is a conjecture that  is meaningful from the point of view of effective theories.  Effective theories are in general non renormalizable. Higher orders would require counter terms with unknown constants to be fitted by experiments. If their philosophy is accepted it corresponds Nature to prove it or disprove it.

\begin{figure}[htb]
\begin{center}
\includegraphics[scale= 0.6]{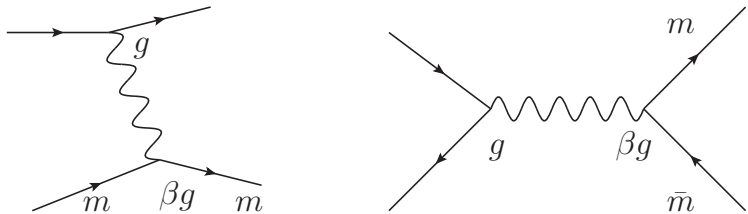}
\caption{ Elementary vertices of monopole-monopole and monopole-antimonopole photon coupling.}
\label{vertices} 
\end{center}
\end{figure}

The Ginzburg-Schiller prescription is only valid below the monopole production threshold, and as such not very powerful from the point of view of phenomenology, because virtual particles, very far off-shell, can be easily confused with other scenarios. 
A complementary idea was proposed in ref. \cite{Kalbfleisch:2000iz} and is beautifully explained in the thesis of Mulhearn \cite{Mulhearn:2004kw}. These authors realized that a monopole interacts with an electron like a (duality transformed) positron, and therefore the effective coupling for electron-monopole is simply $\beta g$, where $\beta$ is the velocity of the monopole. In order to study monopole-antimonopole production processes (via  Drell-Yan) they simply turned the diagram around, thus constructing an effective vertex with coupling $\beta g$ (see Fig.~\ref{vertices}).

%%%%%%%%%%%%%%%%%%%%%%%%%%%%%%%%%%%%%%%%%%%%%%%%%%%%%%%%
%          Fig.1  elementary vertices
%%%%%%%%%%%%%%%%%%%%%%%%%%%%%%%%%%%%%%%%%%%%%%%%%%%%%%%%
%%%%%%%%%%%%%%%%%%%%%%%%%%%%%%%%%%%%%%%%%%%%%%%%%%%%%%%%%%%%%%%

\begin{figure}[htb]
\begin{center}
\includegraphics[scale= 0.6]{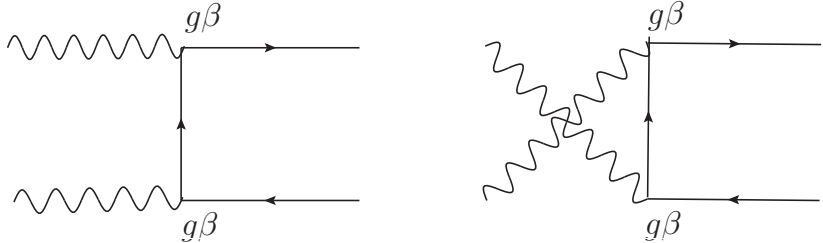}
\caption{ Elementary processes of monopole-antimonopole production via photon fusion.}
\label{mmproduction} 
\end{center}
\end{figure}

This scheme has been used to define an effective field theory  to lowest order \cite{Kalbfleisch:2000iz} which has been applied  to Drell-Yan like monopole-antimonopole production  \cite{Kalbfleisch:2000iz,Abulencia:2005kq} and to monopole-antimonopole production by photon fusion \cite{Kurochkin:2006jr,Dougall:2007tt} (see Fig.~\ref{mmproduction}). 

%%%%%%%%%%%%%%%%%%%%%%%%%%%%%%%%%%%%%%%%%%%%%%%%%%%%%%%%
%          Fig.2 mmproduction by photon fusion
%%%%%%%%%%%%%%%%%%%%%%%%%%%%%%%%%%%%%%%%%%%%%%%%%%%%%%%%
%%%%%%%%%%%%%%%%%%%%%%%%%%%%%%%%%%%%%%%%%%%%%%%%%%%%%%%%%%%%%%%

Non perturbative solutions of field theories lead to momentum-dependent coupling constants and even momentum dependent masses. For example, the resolution of truncated Dyson-Schwinger equations leads to the freezing of the QCD running coupling (effective charge) in the infrared, which is best understood as a dynamical generation of a gluon mass function, giving rise to a momentum dependence which is free from infrared divergences \cite{Cornwall:1982zr}. These studies have motivated our starting hypothesis. We consider the diagrams of the Fig.~\ref{vertices}
as the diagrams of an effective theory where 

\begin{equation}
\alpha (q^2) \sim \beta (q^2) g,
\end{equation}
and moreover, we assume, with Ginzburg and Schiller, that this effective theory is valid only to lowest non-vanishing order.  The momentum dependence is probably much more involved, but we know of two instances where it reduces to $\beta$g, namely those of the figures. Effective theories, as it was said, are not renormalizable, and moreover higher orders in the field expansion will require additional counter terms and new constants to be fitted to the data, therefore our theory is at present only defined to lowest non-vanishing order. To construct the higher order approximation  we should apply Weinberg's theorem \cite{Weinberg:1978kz} and construct all terms compatible with the symmetries. At present, and close to monopole-antimonopole threshold, we expect the lowest order term to be sufficient, for our purposes. Guided by simplicity and phenomenological inspiration we introduce an effective theory which is finite and well defined and we call this proposal the {\emph {$\beta$ scheme}}.

Note that the Ginzburg-Schiller scheme and the $\beta$ scheme are in some sense complementary. The former is valid below the monopole threshold, while the latter above since $\beta$ vanishes below threshold.

The aim here is to study   possible  signals of magnetic monopoles at LHC. According to previous studies \cite{Dougall:2007tt}, the most promising mechanism is photon fusion. The elementary diagrams contributing to pair production are those in Fig.~\ref{mmproduction}, where the explicit couplings have been shown.  

\begin{figure}[htb]
\begin{center}
\includegraphics[scale= 0.9]{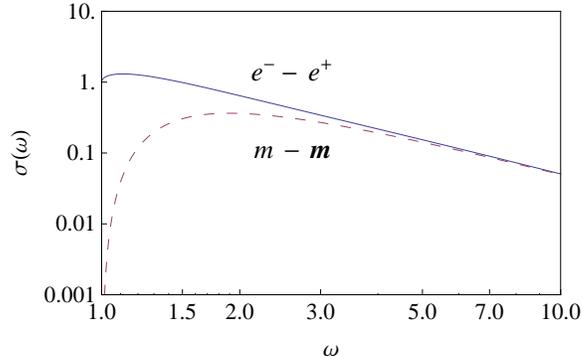}
\caption{ Adimensional functional form of the elementary photon-fusion cross section for electron-positron (solid) and that of the monopole-antimonopole (dashed) as a function of $\omega$ which shows the effect of the $\beta\, g$ coupling. } 
\label{mmxsection1} 
\end{center}
\end{figure}

The photon-fusion elementary cross section  is obtained from the well-known QED electron-positron pair creation cross section \cite{Itzykson:1980rh}, simply changing the coupling constant  ($e \rightarrow g\beta$ ) and the electron mass by the monopole mass $m_e \rightarrow m$, leading to

\begin{equation}
\sigma(\gamma \,\gamma \rightarrow m \overline{m})= \frac{\pi\;g^4
\,(1-\beta^2)\,\beta^4}{2 \,m^2}
\left(\frac{3-\beta^4}{2 \beta}
\log\left({\frac{1+\beta}{1-\beta}}\right) -(2-\beta^2)\right),
\label{ggxsecmm}
\end{equation}
where  $\beta$ is the monopole velocity, a function of the center-of-mass energy, $E$.  In Fig.~\ref{mmxsection1} we show the $\omega= E/2\, m$ dependence of the adimensional functional form of Eq.(\ref{ggxsecmm}) to show the effect of the $\beta\,g$ coupling. The solid curve corresponds to the electron-positron case, the dashed one to the monopole case which contains the $\beta^4$ factor. One should notice the large effect associated with this factor in the vicinity of the threshold.

%%%%%%%%%%%%%%%%%%%%%%%%%%%%%%%%%%%%%%%%%%%%%%%%%%%%%%%%
%          Fig.3 mm production cross section
%%%%%%%%%%%%%%%%%%%%%%%%%%%%%%%%%%%%%%%%%%%%%%%%%%%%%%%%
%%%%%%%%%%%%%%%%%%%%%%%%%%%%%%%%%%%%%%%%%%%%%%%%%%%%%%%%%%%%%%%

LHC detectors, apart from the MoEDAL experiment \cite{moedal},  have not been designed specifically to see monopoles directly and therefore even those which do not annihilate inside the detectors will be difficult to detect. However, the extreme sensitivity of LHC detectors to photons,  due to the importance of the  2$\gamma$ channel in detecting a low mass Higgs, makes them ideal for the purpose of detecting monopole-antimonopole annihilation.

 The two photon process through monopole loops in the Ginzburg-Schiller scheme was studied already sometime ago both theoretically in ref. \cite{Ginzburg:1998vb} and experimentally  in ref. \cite{Abbott:1998mw}. We have proceeded in this paper to perform the calculation in the $\beta$ scheme.

\section{Monopole-antimonopole annihilation into $\gamma \, \gamma$}\label{monopole}

It is natural to think that the enormous strength and long range of the  monopole interaction leads to the annihilation of the pair  into photons very close to the production point. Thus one should  look for monopoles through their annihilation into highly energetic photons, a channel for which LHC detectors have been optimized.
 
\begin{figure}[htb]
\begin{center}
\includegraphics[scale= 0.6]{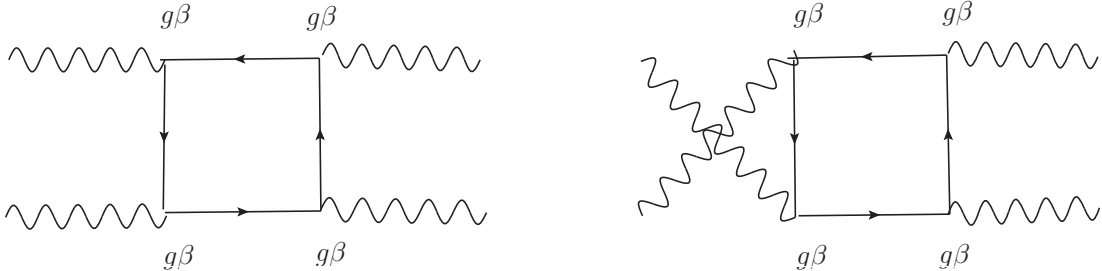}
\caption{ Elementary processes for  monopole-antimonopole production and annihilation into photons.}
\label{mmannihilation} 
\end{center}
\end{figure}

In order to calculate the annihilation into photons we  assume that our effective theory, a technically convenient modification of Ginzburg and Schiller's, agrees with QED at one loop order,  and therefore we apply light-by-light scattering with the appropriate modifications as shown in Fig.~\ref{mmannihilation}.
An interesting feature of the calculation is that the additional magnetic coupling will increase the cross section dramatically and therefore this measurement
should lead to a strong restriction on the monopole mass. However, as we have seen, in $m-\overline{m}$ production, the large magnetic coupling is always multiplied by the small electric one, leading to effective couplings  $e\, g$, and the same will happen in detection. Thus, the effective coupling of the process is $e \, g$, and therefore has strengths similar to the strong interaction, not more. 

%%%%%%%%%%%%%%%%%%%%%%%%%%%%%%%%%%%%%%%%%%%%%%%%%%%%%%%%
%          Fig.4 mmannihilation into photons
%%%%%%%%%%%%%%%%%%%%%%%%%%%%%%%%%%%%%%%%%%%%%%%%%%%%%%%%
%%%%%%%%%%%%%%%%%%%%%%%%%%%%%%%%%%%%%%%%%%%%%%%%%%%%%%%%%%%%%%%

Light-by-light scattering was studied  by Karplus and Neuman \cite{Karplus:1950zza,Karplus:1950zz}, who reproduced all previous low energy results by Euler \cite{Euler:1936zz} and high energy results by Achieser \cite{Achieser:1937zz}, and was later revisited and corrected by Csonka and Koelbig \cite{Csonka:1974ey}.

\begin{figure}[htb]
\begin{center}
\includegraphics[scale= 0.9]{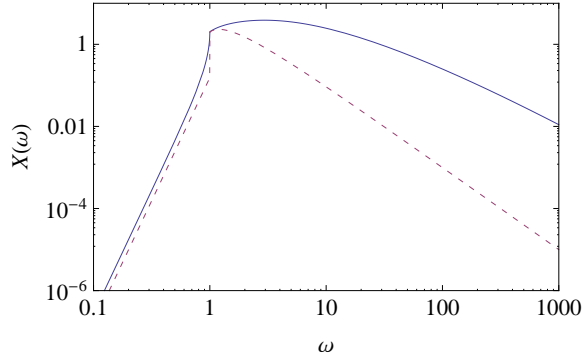}
\caption{The light-by-light scattering quantity $X(\omega)$, which determines the cross section up to factors related to the coupling constant. 
The solid curve corresponds to forward scattering, the dashed one to right-angle scattering, both as a function of the dimensionless quantity $\omega$.}
\label{lightlight} 
\end{center}
\end{figure}

\begin{figure}[htb]
\begin{center}   

\includegraphics[scale= 0.65]{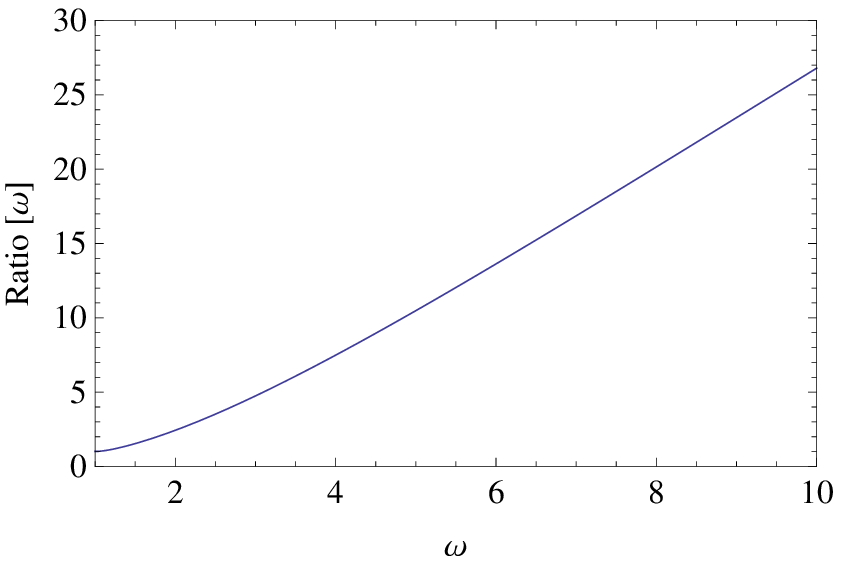}
\hskip 1cm
\includegraphics[scale= 0.65]{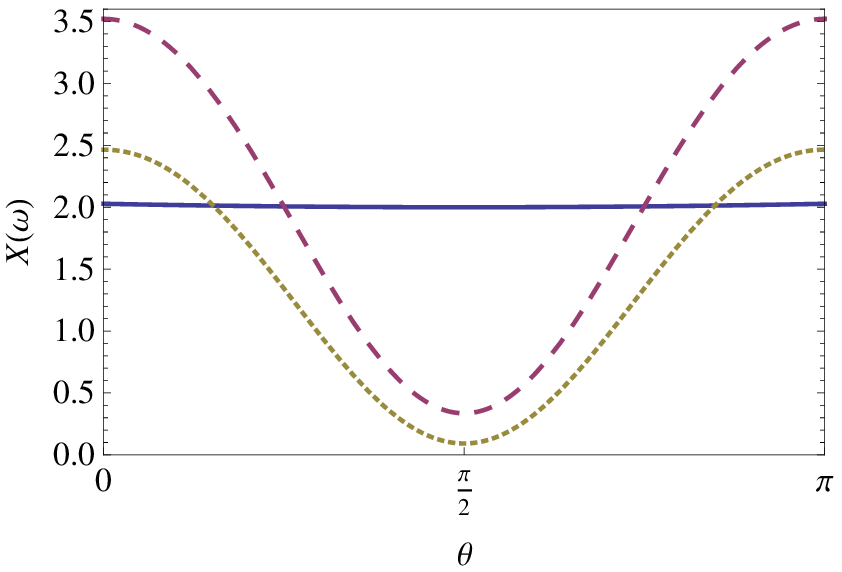}
\caption{Left: ratio of the  light-by-light scattering quantity $X(\omega)$ for forward to right-angle scattering as a function of $\omega$.  Right: angular dependence of the cross section for three values of $\omega$ (1 (solid) , 5 (dashed), 10 (dotted)) . }
\label{comparison} 
\end{center}
\end{figure}

In the case of monopoles, with the appropriate changes, the expression for the cross section becomes, 

\begin{equation}
 \sigma_{\gamma \gamma} (\theta, \omega) = \frac{(g \beta)^8}{8 \pi^2 m^2}  \; X(\theta , \omega).
\label{lightlightxsection}
\end{equation}
where $X(\omega)$ is $\frac{<|M|^2>}{\omega^2}$ in the notation of ref. \cite{Karplus:1950zz}.
In Fig.~\ref{lightlight} we reproduce these results for forward scattering and right-angle scattering.  In Fig.~\ref{comparison} (left)  we show the ratio of the forward scattering  to the right-angle  cross sections and therefore  show how  the cross section, which is basically isotropic close to  threshold, becomes anisotropic as the energy increases. In the same figure (right) we plot  $X(\omega)$ for several values of $\omega$  as a function of angle. We note that the forward cross section is larger and the right-angle one  smaller, than that for any other value of the scattering angle. As the energy increases the drop in the $X$ function from $\theta= 0$ to $\pi/2$ increases. 

%%%%%%%%%%%%%%%%%%%%%%%%%%%%%%%%%%%%%%%%%%%%%%%%%%%%%%%%
%          Fig.5 lightlight scattering
%%%%%%%%%%%%%%%%%%%%%%%%%%%%%%%%%%%%%%%%%%%%%%%%%%%%%%%%
%%%%%%%%%%%%%%%%%%%%%%%%%%%%%%%%%%%%%%%%%%%%%%%%%%%%%%%%%%%%%%%

%%%%%%%%%%%%%%%%%%%%%%%%%%%%%%%%%%%%%%%%%%%%%%%%%%%%%%%%
%          Fig.6   ratio0pi2   and angular dependence
%%%%%%%%%%%%%%%%%%%%%%%%%%%%%%%%%%%%%%%%%%%%%%%%%%%%%%%%
%%%%%%%%%%%%%%%%%%%%%%%%%%%%%%%%%%%%%%%%%%%%%%%%%%%%%%%%%%%%%%%

 It is clear from these figures that close to the threshold the cross section is quite isotropic and  away from threshold the forward cross section, which is very difficult to measure, is much larger than the right-angle one.  Since the detectors cannot detect all of the photons coming out, we  take the right-angle cross section as a conservative indication of the magnitudes to be expected.

\begin{figure}[htb]
\begin{center}   
\includegraphics[scale=0.9]{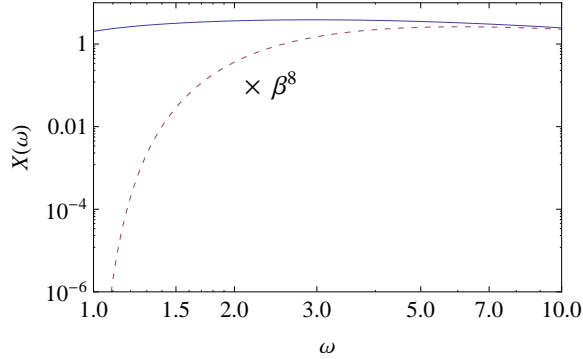}
\caption{The effect of the $\beta$ factor  as a function of $\omega$. The solid curve corresponds to electron scattering, while the dashed curve corresponds to 
the monopole case. }
\label{beta} 
\end{center}
\end{figure}

In Fig.~\ref{beta} we show the effect of the $\beta$ factor which diminishes greatly the cross section close to the monopole-antimonopole threshold.  

%%%%%%%%%%%%%%%%%%%%%%%%%%%%%%%%%%%%%%%%%%%%%%%%%%%%%%%%
%          Fig.7   beta effect
%%%%%%%%%%%%%%%%%%%%%%%%%%%%%%%%%%%%%%%%%%%%%%%%%%%%%%%%
%%%%%%%%%%%%%%%%%%%%%%%%%%%%%%%%%%%%%%%%%%%%%%%%%%%%%%%%%%%%%%%

\section{Monopolium annihilation into $\gamma \; \gamma$}\label{monopolium}

\begin{figure}[htb]
\centerline{\epsfig{file=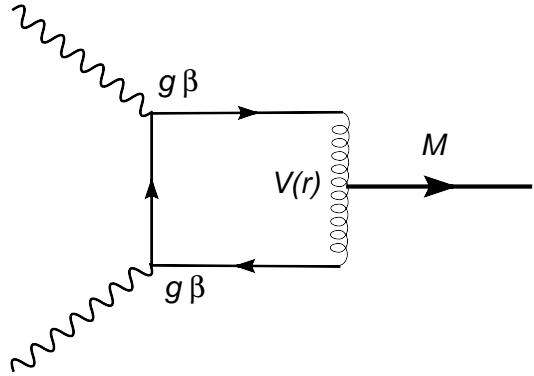,width=7cm,angle=0}}
\caption{\small{Diagrammatic description of the elementary
subprocess of the monopolium production from photon fusion. 
$V(r)$ represents the interaction binding the monopole-antimonopole pair to form monopolium.}}
\label{gbetab}\end{figure}

Recently we studied the production of monopolium by photon 
fusion at LHC \cite{Epele:2008un}. 
The elementary subprocess calculated is shown in Fig.~\ref{gbetab}. The
standard expression for the cross section of the elementary
subprocess for producing a monopolium of mass $M$ is given by

\begin{equation}
\sigma (2 \gamma \rightarrow M) = \frac{4\pi}{E^2}  \frac{M ^2
\,\Gamma (E) \, \Gamma_M}{\left(E^2 - M^2\right)^2 +
M^2\,\Gamma_M^2}, \label{ppM}
\end{equation}
where  $\Gamma (E)$, with $E$ off mass shell, describes the production
cross section. Note that $\Gamma[M] = 0$. $\Gamma_M$ arises from the softening of the
delta function, $\delta(E^2 - M^2)$ and therefore is, in principle,
independent of the production rate $\Gamma (E)$ and can be attributed to the beam 
width \cite{Jauch:1975sp,Peskin:1995hc}. 

%%%%%%%%%%%%%%%%%%%%%%%%%%%%%%%%%%%%%%%%%%%%%%%%%%%%%%%%
%          Fig.8 Mproduction by photon fusion 
%%%%%%%%%%%%%%%%%%%%%%%%%%%%%%%%%%%%%%%%%%%%%%%%%%%%%%%%
%%%%%%%%%%%%%%%%%%%%%%%%%%%%%%%%%%%%%%%%%%%%%%%%%%%%%%%%%%%%%%%

\begin{figure}[htb]
\centerline{\epsfig{file=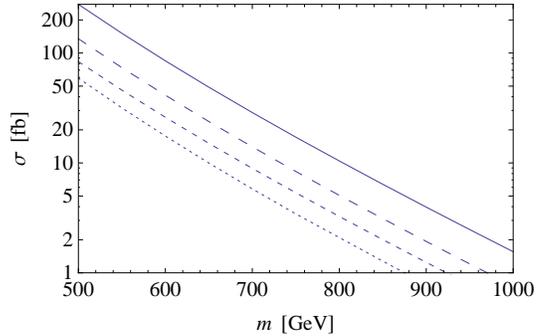,width=7cm,angle=0}}
\caption{\small{Total cross section for monopolium production at LHC with $3.5$ TeV beams for monopole masses ranging from $500$ to $1000$ GeV (full curve). The broken curves represent the different contributions to the total cross section as described in the text: semi-elastic (dashed), elastic (short dashed) and inelastic (dotted). We have chosen 
a binding energy $\sim 2\;m/15$ and  $\Gamma_M = 10$ GeV. }}
\label{Mproduction}\end{figure}

In Fig.~\ref{Mproduction} we show the total cross section for monopolium production from photon fusion
under present LHC running conditions for a monopole mass ($m$)  ranging from $500$ to $1000$ GeV. 
In the figure the binding energy is fixed for each mass  ($2\;m/15$), chosen so that for our case study, $m=750$ GeV, the binding energy is
$100$ GeV and thus $M=1400$ GeV. With this choice the monopolium mass ($M$) ranges from $933$ to $1866$ GeV. 
We notice that detection would be possible with an integrated luminosity of $5$ fb$^{-1}$ if the chosen binding energy
is at the level of $10\% $ of the monopole mass or higher. In the present analysis we study  binding energies small compared to the bound state mass, $M$, in order to be consistent  with the formalism 
used.

%%%%%%%%%%%%%%%%%%%%%%%%%%%%%%%%%%%%%%%%%%%%%%%%%%%%%%%%
%          Fig.9 Mproduction by photon fusion 
%%%%%%%%%%%%%%%%%%%%%%%%%%%%%%%%%%%%%%%%%%%%%%%%%%%%%%%%
%%%%%%%%%%%%%%%%%%%%%%%%%%%%%%%%%%%%%%%%%%%%%%%%%%%%%%%%%%%%%%%

\begin{figure}[htb]
\centerline{\epsfig{file=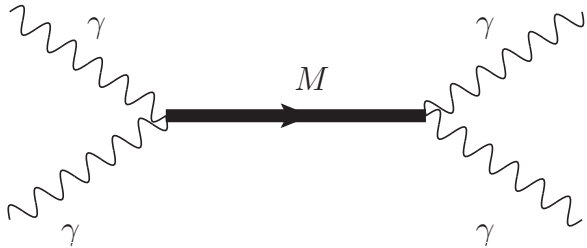,width=8cm,angle=0}}
\caption{\small{Diagrammatic description of the monopolium production and decay.}}
\label{ggMgg}\end{figure}

The interest in this paper is in the detection of photons after monopolium decay. 
The elementary subprocess is shown in Fig.~\ref{ggMgg}, which could be considered as a contribution to light-by-light scattering
in the presence of  monopolium.

%%%%%%%%%%%%%%%%%%%%%%%%%%%%%%%%%%%%%%%%%%%%%%%%%%%%%%%%
%          Fig.10 gamma gamma M gamma gamma 
%%%%%%%%%%%%%%%%%%%%%%%%%%%%%%%%%%%%%%%%%%%%%%%%%%%%%%%%
%%%%%%%%%%%%%%%%%%%%%%%%%%%%%%%%%%%%%%%%%%%%%%%%%%%%%%%%%%%%%%%

The standard expression for the cross section of this elementary
subprocess, after having integrated over angles, is given by 

\begin{equation}
\sigma (\gamma\, \gamma \rightarrow M \rightarrow \gamma \, \gamma ) = \frac{4\pi}{E^2}  \frac{M ^2
\,\Gamma^2 (E)}{\left(E^2 - M^2\right)^2 +
M^2\,\Gamma_M^2}. \label{ppM2}
\end{equation}
Here  $\Gamma (E)$, 
with $E$ off mass shell, describes the vertex $\gamma\, \gamma \, M$.
Monopolium is stable in the center of mass but we add an experimental Gaussian  width $\Gamma_M \sim 10$ 
GeV  in line with the values used in ref. \cite{Allanach:2000nr}. 

\begin{figure}[htb]
\centerline{\epsfig{file=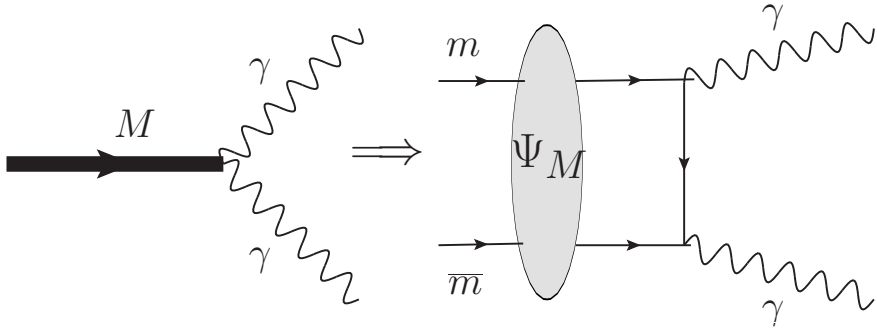,width=12cm,angle=0}}
\caption{\small{The monopolium vertex and its microscopic description
.}}
\label{Mggvertex}\end{figure}

We recall now the computation of the $\Gamma (E)$, which represents
the vertex of the monopolium decay to $\gamma\, \gamma$ . The calculation,
following standard field-theory techniques of the decay of a
non-relativistic bound state,
leads to

\begin{equation}
\Gamma (E) =
\frac{32\,\pi\,\alpha_g^2}{M^2}\,\left|\psi_M(0)\right|^2 .
\end{equation}
We have used the conventional approximations for this calculation:  the monopole and antimonopole,  
forming the bound state, are treated as on-shell particles,  when calculating the elementary scattering process shown
on the right of Fig.~\ref{Mggvertex}; the bound state is described by a wave function obtained from a Coulomb-type 
interaction between the pair \cite{Epele:2008un,Jauch:1975sp,Peskin:1995hc}. However, 
once the calculation is performed  we substitute $2m$ by $M$, where $m$ is the monopole
 mass  to take into account  binding. In the expression, 
 $\alpha_g$ corresponds to the photon--monopole
coupling and $\psi_M$ is the monopolium ground state wave function.

%%%%%%%%%%%%%%%%%%%%%%%%%%%%%%%%%%%%%%%%%%%%%%%%%%%%%%%%
%          Fig.11 M gamma gamma vertex
%%%%%%%%%%%%%%%%%%%%%%%%%%%%%%%%%%%%%%%%%%%%%%%%%%%%%%%%
%%%%%%%%%%%%%%%%%%%%%%%%%%%%%%%%%%%%%%%%%%%%%%%%%%%%%%%%%%%%%%%

Before we proceed let us raise a warning on our calculation. Bound 
relativistic systems are notoriously difficult dynamical objects. We proceed here
by performing a non-relativistic calculation of the monopolium wave function. 
The validity of the non-relativistic approximation in the bound state wave functions 
was analyzed in a previous work \cite{Epele:2007ic}. In the dynamics of the decay 
formula the substitution of $2m$ by $M$ constitutes an intuitive way 
to take the off-shellness of the monopoles, i.e. their binding energy in monopolium,  into account. 
For the purposes of estimation both approximations seem reasonable specially
since our binding energies will never exceed $15\%$ of the monopole mass, i.e. less than $10\%$ 
of the total mass of the system.

Using the Coulomb wave functions of ref.\cite{Epele:2007ic}
expressed in the most convenient way to avoid details of the
interaction, which will be parameterized by the binding energy, one
has

\begin{equation}
|\psi_{M}(0)|^2 = \frac{1}{\pi}\left(2 - \frac{M}{m}\right)^{3/2}\; m^3,
\end{equation}
and the effective monopole coupling theory described above 
in the case of monopolium
production, gives rise to \cite{Epele:2008un},

\begin{equation}
\frac{\Gamma(E)}{M}= 2\left(\frac{ \beta^2}{\alpha}\right)^2 \left(\frac{m}{M}\right)^3  \left(2- \frac{M}{m}\right)^{3/2}.
\end{equation}
Here, $\alpha$ is the fine structure constant and $\beta$ the
monopolium velocity,

\begin{equation}
\beta= \sqrt{1-\frac{M^2}{E^2}} ,
\end{equation}
which is the velocity of the monopoles moving in the monopolium
system.

Note that due to the value of $\beta$ the vertex vanishes at the
monopolium mass, where the velocity is zero. Therefore a static
monopolium is stable under this interaction. We refer to refs. 
\cite{Mulhearn:2004kw,Kalbfleisch:2000iz} for a thorough discussion 
on Lorentz invariance of the theory.

A caveat is due here. There is a duality of treatments in the above
formulation as can be seen in Fig.~\ref{gbetab}. The static coupling is treated
as a Coloumb like interaction of coupling $g$ binding the monopoles
into monopolium, although ultimately the details are eliminated in
favor of the binding energy parameterized by the monopolium mass
$M$. We find in this way a simple parametric description of the
bound state. The dynamics of the production of the virtual
monopoles, to be bound in monopolium, is described in accordance
with the effective theory \cite{Kalbfleisch:2000iz}, and
this coupling is $\beta g$. This is similar to what is done in heavy
quark physics \cite{Pennington:2005ww}(see his figure 5), where the
wave function is obtained by a parametric description using
approximate strong dynamics while the coupling to photons is
elementary.

The production cross section can now be written as,

\begin{equation}
M^2\;\sigma(\gamma\,\gamma \rightarrow M\rightarrow\gamma\,\gamma ) = 16 \;\pi
 \left(\frac{\beta}{\alpha}\right)^4  \left(\frac{m}{E}\right)^6\,\left(2 - \frac{M}{m}\right)^3\,\left(1 + \frac{M^2 \Gamma_M^2}{E^4 \beta^4}\right)^{-1}.
\label{ggxsecM}\end{equation}

The above cross 
section satisfies comfortably the unitarity limit 
\cite{Milton:2008pn},
\begin{equation}
\sigma \le \frac{\pi}{3 E^2}.
\end{equation}

\begin{figure}[htb]
\epsfig{file=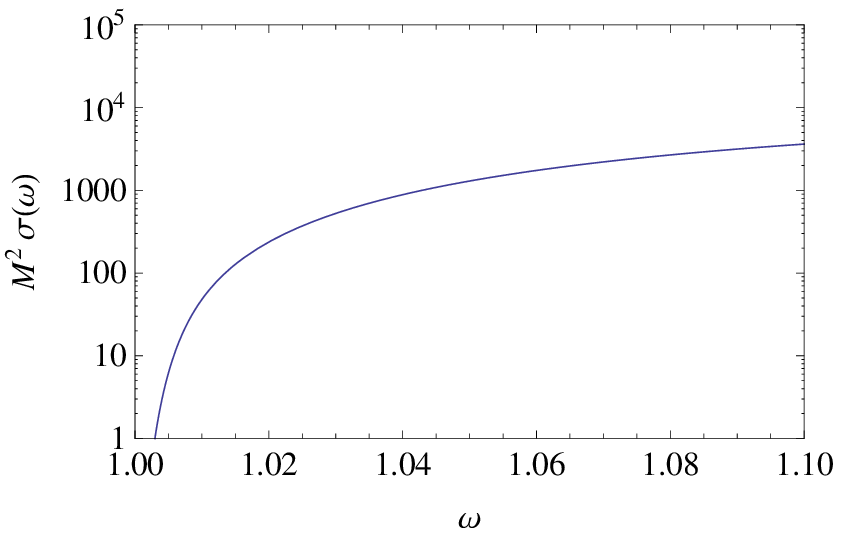,width=6.5cm,angle=0} \hspace{1cm}
\epsfig{file=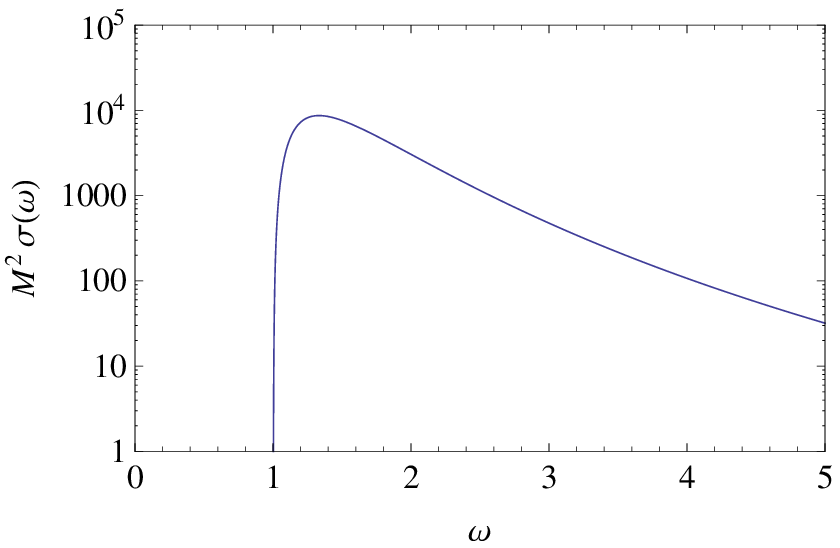,width=6.5cm,angle=0}
\caption{\small{Elementary cross section as a function of $\omega = E/M$ in units of $1/M^2$ calculated for $m=750$ GeV and $M= 1400$ GeV. Left: near threshold. Right:  away from threshold.}} \label{xsec}\end{figure}

To feel safe with our approximations we consider the binding energy much smaller than $m$, i.e. $M \sim 2m$. In this case the elementary cross section has two very different behaviors as shown in  Fig.~\ref{xsec} : i) at threshold it is dominated by $\beta$ and the cross section rises (see left figure); ii) away from threshold the dominant behavior is the $1/E$ dependence and the cross section drops faster than the unitary limit. The conflict between these two behaviors produces a  wide bump-like structure.

%%%%%%%%%%%%%%%%%%%%%%%%%%%%%%%%%%%%%%%%%%%%%%%%%%%%%%%%
%          Fig.12  elementary sigma
%%%%%%%%%%%%%%%%%%%%%%%%%%%%%%%%%%%%%%%%%%%%%%%%%%%%%%%%
%%%%%%%%%%%%%%%%%%%%%%%%%%%%%%%%%%%%%%%%%%%%%%%%%%%%%%%%%%%%%%%

\section{Analysis of $p-p$ scattering}\label{scattering}

\begin{figure}[htb]
\begin{center}   
\includegraphics[scale=0.5]{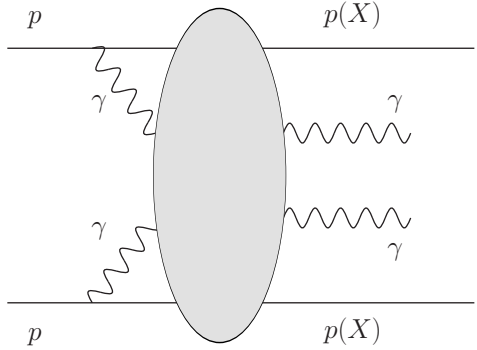}
\caption{Processes contributing to the $\gamma \gamma$ cross section. The blob contains the three cases described in the text. }
\label{pp} 
\end{center}
\end{figure}

LHC is a proton-proton collider, therefore, in order to describe the production and desintegration of the monopole-antimonopole pair, 
we have to study the following processes above the monopole threshold ($ \beta > 0 \rightarrow E \geq 2 m$),

\begin{eqnarray}
p + p &  \rightarrow &  p(X) + p(X) + \gamma + \gamma,
\end{eqnarray}
shown globally in Fig.\ref{pp}, where $p$ represents the proton, $X$ an
unknown final state and we will assume that the blob is due exclusively to a) monopoles and b) monopolium. This diagram summarizes
the three possible processes in each case:\\[0.2cm] 

\noindent a) monopole-antimonopole annihilation

\begin{itemize}

\item [  i)] inelastic $p+ p \rightarrow X+X + \gamma +\gamma
\rightarrow X + X + m + \overline{m} \rightarrow  X + X + m + \overline{m} +\gamma + \gamma$

\item [ ii)] semi-elastic $ p + p \rightarrow p + X + \gamma + \gamma
\rightarrow p + X +  m + \overline{m}  \rightarrow  p + X + m + \overline{m} +\gamma + \gamma$

\item [iii)] elastic $p + p \rightarrow p + p + \gamma + \gamma
\rightarrow p + p +  m + \overline{m} \rightarrow \rightarrow  X + X + m + \overline{m} + \gamma + \gamma$.
\end{itemize}
\vskip 0.1cm
and\\[0.2cm]

\noindent b) monopolium annhilation
\begin{itemize}

\item [i)] inelastic $p+ p \rightarrow X+X + \gamma +\gamma
\rightarrow X + X +M \rightarrow  X + X + M +\gamma + \gamma$

\item [ii)] semi-elastic $ p + p \rightarrow p + X + \gamma + \gamma
\rightarrow p + X + M \rightarrow  p + X + M +\gamma + \gamma$

\item [iii)] elastic $p + p \rightarrow p + p + \gamma + \gamma
\rightarrow p + p +  M \rightarrow  p + p  + M + \gamma + \gamma$.
\end{itemize}

%%%%%%%%%%%%%%%%%%%%%%%%%%%%%%%%%%%%%%%%%%%%%%%%%%%%%%%%
%          Fig.13   p p scattering
%%%%%%%%%%%%%%%%%%%%%%%%%%%%%%%%%%%%%%%%%%%%%%%%%%%%%%%%
%%%%%%%%%%%%%%%%%%%%%%%%%%%%%%%%%%%%%%%%%%%%%%%%%%%%%%%%%%%%%%%

In the inelastic scattering, both intermediate photons are radiated
from partons (quarks or  antiquarks) in the colliding protons.

In the semi-elastic scattering one intermediate photon is radiated
by a quark (or antiquark), as in the inelastic process, while the
second photon is radiated from the other proton, coupling to the
total proton charge and leaving a final state proton intact.

In the elastic scattering both intermediate photons are radiated
from the interacting protons leaving both protons intact in the
final state.

In the blob we incorporate in a) the elementary subprocess shown in Fig.~\ref{mmannihilation} and  described by Eq. (\ref{lightlightxsection}) and in b) the elementary subprocess shown in Fig.~\ref{ggMgg} and  described by Eq. (\ref{ggxsecM}).

We calculate $\gamma \gamma$ fusion for monopole-antimonopole and monopolium production
following the formalism of Drees et al. \cite{Drees:1994zx}.

In the inelastic scattering, $p + p\rightarrow X+ X + (\gamma
\gamma) \rightarrow X +X + (m+ \overline{m}) ( \mbox{or} \;M)  + \gamma + \gamma$, to approximate the quark distribution
within the proton we use the Cteq6--1L parton distribution functions
\cite{CTEQ} and choose $Q^2=\hat{s}/4$ throughout, where $\hat{s}$ is the center of
mass energy of the elementary process.

We employ an equivalent--photon approximation for the photon
spectrum of the intermediate quarks \cite{Williams:1934ad,vonWeizsacker:1934sx}.

In semi-elastic scattering, $p + p\rightarrow p+ X+ (\gamma \gamma)
\rightarrow p+ X + (m+ \overline{m}) (\mbox{or} \; M)  + \gamma + \gamma $, the photon spectrum associated with the
interacting proton must be altered from the equivalent--photon
approximation for quarks to account for the proton structure.  To
accommodate the proton structure we use the modified
equivalent--photon approximation of \cite{Drees:1994zx}.

The total cross section is obtained 
as a sum of the three processes. The explicit expressions for the
different contributions can be found  in \cite{Dougall:2007tt}.

In order to obtain the differential photon-photon cross section from the above
formalism we develop a procedure  which we exemplify with the elastic 
scattering case. In that case the $pp$ cross section is given by \cite{Drees:1988pp,Drees:1994zx},

\begin{equation}
\sigma_{pp} (s)  = \int_{s_{th}/s}^1 d z_1  \int_{s_{th}/s z_1}^1 d z_2 f(z_1) f(z_2) \sigma_{\gamma \gamma} (z_1 z_2  s),
\end{equation}
where  $\sqrt{s_{th}} = 2 m\; (\mbox{or}\; M)$ is the threshold center of mass energy, $\sqrt{s}$ is the center of mass energy of the $pp$ system  and the $f$'s represent the elastic photon spectrum.

We perform the following change of variables 
$$ v= z_1 z_2  \;  , \; w= z_2 \; ,$$
which leads to 
\begin{equation}
\sigma_{pp} (s)  = \int_{s_{th}/s}^1 d v  \int_{v}^1 \frac{d w}{w}  f(\frac{v}{w}) f(w) \sigma_{\gamma \gamma} (v s). 
\end{equation}
Note that to fix the center of mass energy of the photons is equivalent to fix $v$. For fixed $v$ we have,

\begin{equation}
\frac{d \sigma_{pp}}{dv} (s)  = \int_{v}^1  \frac{d w}{w}  f(\frac{v}{ w}) f(w) \sigma_{\gamma \gamma} (v s),
\end{equation}
which can be rewritten in terms  of $E_\gamma$, the center of mass energy  of the photons, and the elementary photon-photon cross section as,

\begin{equation}
 \frac{d \sigma_{pp}}{dE} (E_\gamma) = \frac{2 E_\gamma}{s}  \; \sigma_{\gamma \gamma} (s_{\gamma \gamma})\;\int_{s_{\gamma  \gamma}/s}^1  \frac{d w}{w}  f(\frac{s_{\gamma  \gamma}}{ w}) f(w).  
\end{equation}
This procedure can be generalized easily to the semi-elastic and inelastic cases, where the appropriate change of variables are

$$ v= z_1 z_2 x_1  \; , \;  w= z_2 x_1 \; , \;  u =  x_1$$
and
$$ v= z_1 z_2 x_1  x_2 \; , \;  w= z_2 x_1 x_2 \; , \;  u =  x_1 x_2 \; , \; t = x_2,$$
respectively which one has to introduce into integral expressions with  a product of three $f$'s (semi-elastic) or four $f$'s (inelastic) representing  quark densities and photon spectrum \cite{Drees:1994zx}.

%%%%%%%%%%%%%%%%%%%%%%%%%%%%%%%%%%%%%%%%%%%%%%%%%%%%%%%%
%          Fig.14  monopole production by  p p scattering
%%%%%%%%%%%%%%%%%%%%%%%%%%%%%%%%%%%%%%%%%%%%%%%%%%%%%%%%
\begin{figure}[htb]
\begin{center}   
\includegraphics[scale=0.8]{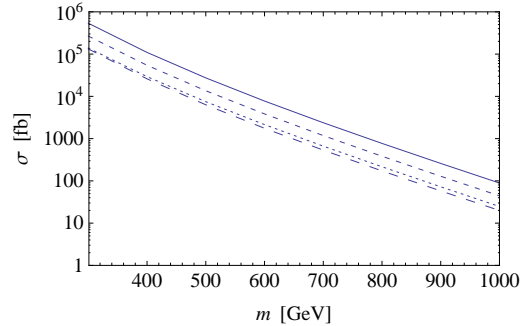}
\caption{The total monopole-antimonopole production cross section (solid line) from $\gamma \, \gamma$ fusion in fb as a function of monopole mass.  The different components of the cross section are shown: elastic (dotted line), semi-elastic (small dashed line) and inelastic (dashed line). }
\label{mmproductionxsec} 
\end{center}
\end{figure}
%%%%%%%%%%%%%%%%%%%%%%%%%%%%%%%%%%%%%%%%%%%%%%%%%%%%%%%%%%%%%%%

\section{Results and Discussion}\label{results}

\subsection{Monopole-Antimonopole Annihilation}

In Fig.~\ref{mmproductionxsec} we show the total cross section for monopole-antimonopole  production and the contribution 
from each of the individual processes described above as a function of monopole mass. In order to set the mass axes we have taken into account the previous 
lower mass limit for the monopole of ref. \cite{Abulencia:2005kq} which was set at $360$ GeV. The cross section is of $\mathcal{O}({\rm fb})$ thus we limit the high mass values to those potentially observable at present, or in the near future, by the LHC with an integrated luminosity of 5~fb$^{-1}$.
 Our result differs from that of ref. \cite{Dougall:2007tt} around threshold.

We see in the previous figure that monopoles of mass around 500 GeV should be produced abundantly, while those of mass around 1000 GeV have a cross section which makes them difficult to detect in the near future, at least directly.

\begin{figure}[htb]
\begin{center}   
\includegraphics[scale=0.8]{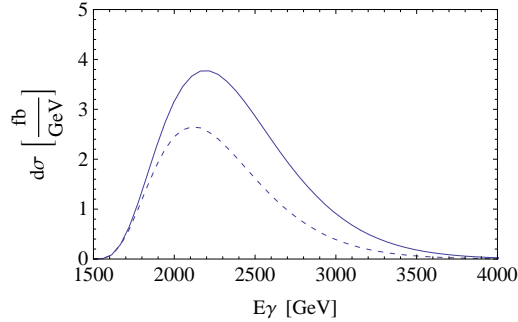}
\caption{Forward (solid) and  right-angle (dashed) cross sections for a monopole of mass $750$ GeV.}
\label{0pi2} 
\end{center}
\end{figure}

Since, as already mentioned, the LHC detectors have not been tuned to detect monopoles but are excellent detectors for $\gamma$'s let us discuss next our annihilation cross section.  In Fig.~\ref{0pi2} we show the differential cross section for forward (solid) and right-angle (dotted) scattering  given in  fb/GeV as a function of the invariant mass of the $\gamma\, \gamma$ system. We have assumed a monopole of mass $750$ GeV, chosen because the cross sections turned out to be close  to the expected magnitude of the  Higgs to $\gamma \; \gamma$ cross section above background. The cross sections are wide, almost gaussian, structures rising softly just above threshold ($1500$ GeV). LHC detectors are blind for forward scattering and have black spots due to construction features in the non-forward regions which do not allow for a full detection of photons. Therefore in order to obtain an educated estimation of  the observable cross section we  take  the right-angle  cross section and multiply it by $4 \pi$. This differential cross section is the smallest possible. However, away from threshold, it corresponds quite well to a realistic estimate, since, as we have seen, the elementary differential cross section drops fast with angle and moreover one should consider an efficiency factor for the various detectors. 

%%%%%%%%%%%%%%%%%%%%%%%%%%%%%%%%%%%%%%%%%%%%%%%%%%%%%%%%%%%%%%%%%%%%%%%%%%%%%%%%%%%%%%%%%%%%%%%%%%%%%%%%%%%
%          Fig. 15  p p scattering for 0 and pi/2
%%%%%%%%%%%%%%%%%%%%%%%%%%%%%%%%%%%%%%%%%%%%%%%%%%%%%%%%
%%%%%%%%%%%%%%%%%%%%%%%%%%%%%%%%%%%%%%%%%%%%%%%%%%%%%%%%%%%%%%%

\begin{figure}[htb]
\begin{center}   
\includegraphics[scale=0.8]{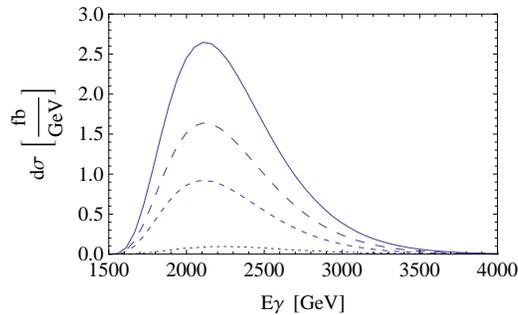}
\caption{The right-angle scattering cross sections for a monopole mass of $750$ GeV.  The smallest is the inelastic cross section (dotted), next comes the elastic (dashed) and the biggest is the semi-elastic (long-dashed). The total cross section, the sum of the three, is represented by the solid curve.}
\label{pi2} 
\end{center}
\end{figure}

 In Fig.~\ref{pi2} we plot the right-angle differential cross section for the same monopole mass, showing the various contributions. The cross section features a wide distribution, rising softly after threshold, $1500$ GeV, and extending for almost $2000$ GeV. The structure is centered about $2300$ GeV. Clearly the soft rise of the differential cross section is a signature of the two-particle threshold, reminiscent of the $\beta$ factor. The width of the structure is associated to the mathematical form of the box diagram, as can be seen for both electron-positron annihilation and monopole-antimonopole, from the form of the elementary cross section (Fig.~\ref{beta}).

%%%%%%%%%%%%%%%%%%%%%%%%%%%%%%%%%%%%%%%%%%%%%%%%%%%%%%%%%%%%%%%%%%%%%%%%%%%%%%%%%%%%%%%%%%%%%%%%%%%%%%%%%%%%%%
%          Fig.16  right angle  scattering
%%%%%%%%%%%%%%%%%%%%%%%%%%%%%%%%%%%%%%%%%%%%%%%%%%%%%%%%
%%%%%%%%%%%%%%%%%%%%%%%%%%%%%%%%%%%%%%%%%%%%%%%%%%%%%%%%%%%%%%%

\begin{figure}[htb]
\begin{center}   
\includegraphics[scale=0.9]{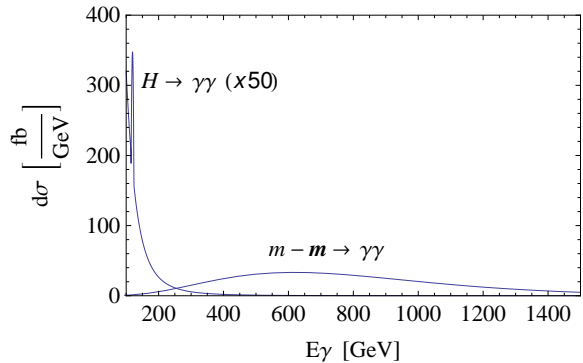}
\caption{Comparison of the $\gamma \gamma$ monopole-antimonopole annihilation cross section for a monopole of mass $750$ GeV  with the Higgs $\gamma \gamma$ decay. The Higgs cross section above the background has been multiplied by $50$. The monopole-antimonopole threshold has been set at the origin ($100$ GeV) (left).}
\label{higgsmonopole} 
\end{center}
\end{figure}

In Fig.~\ref{higgsmonopole} we compare our total $\gamma \gamma$ cross section with that of the Higgs process obtained from ref. \cite{Aad:2010qr}. We have extrapolated their background to our energies using an inverse polynomial fit to their data and their exponential fit. Both procedures give a negligible background for the signals obtained  with monopoles masses up to 1 TeV and even higher. In the left figure we  translate the monopole-antimonopole threshold to the origin in order to compare the two signals. The Higgs signal above background has been multiplied by 50 to make it visible. The figures correspond to a monopole of mass $750$ GeV. It is clear from the curves that monopole-antimonopole annihilation should appear as a soft rise of the cross section  above the background over a large energy interval. Actually the expected background from Standard Model processes is negligible in the kinematic region $E_{\gamma\gamma} \sim 1~{\rm TeV}$. Hence the required selection criteria can be kept minimal, retaining thus the majority of the photon pairs produced in $m-\bar{m}$ annihilation in the analysis. Thus the search for monopole-antimonopole pairs will be practically background free at the LHC.

%%%%%%%%%%%%%%%%%%%%%%%%%%%%%%%%%%%%%%%%%%%%%%%%%%%%%%%%%%%%%%%%%%%%%%%%%%%%%%%%%%%%%%%%%%%%%%%%%%%%%%%%%%%%%
%          Fig. 17   Higgs and monopole -antimonopole to \gamma gamma  
%%%%%%%%%%%%%%%%%%%%%%%%%%%%%%%%%%%%%%%%%%%%%%%%%%%%%%%%
%%%%%%%%%%%%%%%%%%%%%%%%%%%%%%%%%%%%%%%%%%%%%%%%%%%%%%%%%%%%%%%

\begin{figure}[htb]
\begin{center}   
\includegraphics[scale=0.8]{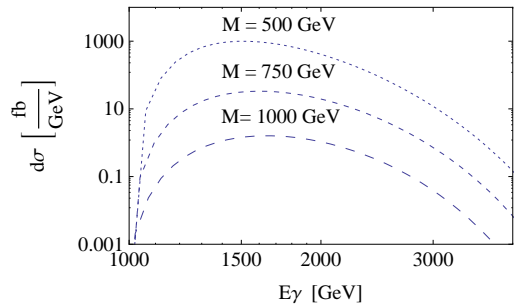}
\caption{The cross section for monopole-antimonopole annihilation into $\gamma \gamma$ for three masses: $500$ GeV (solid), $750$ GeV dashed $1000$ GeV (dotted).}
\label{masses} 
\end{center}
\end{figure}

Finally we study the mass dependence of the differential cross section. Fig.~\ref{masses} is a LogLog plot of the cross section for three masses, $500, 750$ and $1000$ GeV. In order to have a better comparison we have translated the thresholds to the origin. It is clear that the magnitude of the cross section and its extent falls rapidly with mass. For a low mass monopole, the  magnitude of the differential cross section is of the order of  $\sim 1$ pb/GeV, for an intermediate mass monopole the cross section is of the order of   $\sim 10$ fb/GeV  and  for a heavy  monopole is of the order of $\sim 1$ fb/GeV.  If the monopole has a low mass, monopoles should be seen within the initial period of LHC running. 

%%%%%%%%%%%%%%%%%%%%%%%%%%%%%%%%%%%%%%%%%%%%%%%%%%%%%%%%%%%%%%%%%%%%%%%%%%%%%%%%%%%%%%%%%%%%%%%%%%%%%%%%%%%%%
%          Fig. 18  monopole mass discussion
%%%%%%%%%%%%%%%%%%%%%%%%%%%%%%%%%%%%%%%%%%%%%%%%%%%%%%%%
%%%%%%%%%%%%%%%%%%%%%%%%%%%%%%%%%%%%%%%%%%%%%%%%%%%%%%%%

\subsection{Monopolium Annihilation}

\begin{figure}[htb]
\begin{center}   
\includegraphics[scale=0.9]{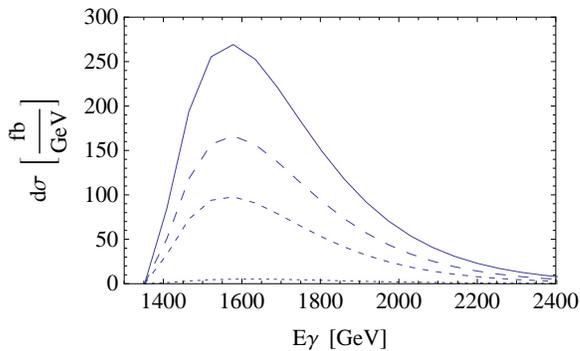}
\caption{The differential cross section (solid line) and its components (semi-elastic (dashed), elastic (short-dashed) and inelastic (dotted)) 
as a function of the center-of-mass photon energy for a monopole mass of $750$ GeV and a monopolium mass of $1400$~GeV. }
\label{ppMggXsec} 
\end{center}
\end{figure}

Our aim is to show scenarios which could arise during the present  LHC running period and to discuss general properties 
of the monopolium system which might serve when higher luminosities are achieved.  In Fig.~\ref{ppMggXsec} we show the 
structure for the differential cross section. It is a wide bump, starting very close after threshold, i.e. the 
monopolium mass ($1400$ GeV in this case), and extending for about $1000$ GeV . We show in the figure the contribution of the different 
components to the cross section.  The elastic and semi-elastic components dominate. The behavior is well understood by the structure of Eq. \ref{ggxsecM}, 
the bump initiates due to the rising of the cross section close to threshold associated with its $\beta$ behavior. Close to threshold  $\beta$ takes
almost its asymptotic value of 1 and the $1/E$ behavior of the cross section starts to softly dominate  (recall Fig.~\ref{xsec}).

%%%%%%%%%%%%%%%%%%%%%%%%%%%%%%%%%%%%%%%%%%%%%%%%%%%%%%%%
%          Fig. 19  M crossection m=750 M=1400
%%%%%%%%%%%%%%%%%%%%%%%%%%%%%%%%%%%%%%%%%%%%%%%%%%%%%%%%
%%%%%%%%%%%%%%%%%%%%%%%%%%%%%%%%%%%%%%%%%%%%%%%%%%%%%%%%

\begin{figure}[htb]
\begin{center}   
\includegraphics[scale=0.9]{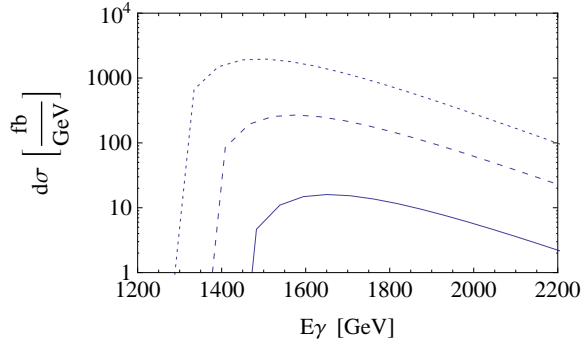}
\caption{Differential two photon cross section as a function of the photon center-of-mass energy for a monopole mass of $750$ GeV 
and different binding energies for monopolium: solid line $75$ GeV, dashed $150$ GeV and dotted $225$ GeV. }
\label{MggEbind} 
\end{center}
\end{figure}

\begin{figure}[htb]
\begin{center}   
\includegraphics[scale=0.9]{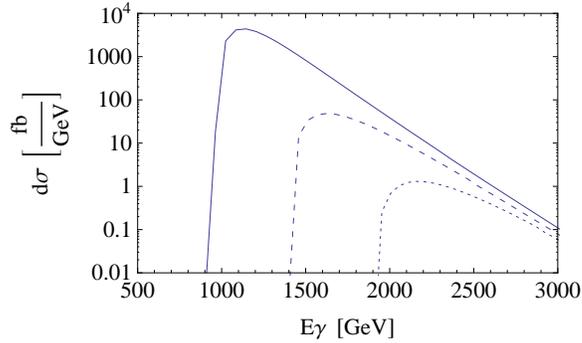}
\caption{Differential cross section as a function of photon energy for different monopole masses: $500$ GeV (solid), $750$ GeV (dashed), $1000$ GeV (dotted) and fixed binding energy ($100$ GeV).}
\label{Mmonopolemasses} 
\end{center}
\end{figure}

Two are the main physical dependences  of the cross section: the monopole mass, $m$,  and the monopolium mass, $M$. In Fig.~\ref{MggEbind} we fix the monopole mass to $750$ GeV and vary the binding energy. We see that the cross section increases dramatically with binding energy. Thus the effect of the binding is twofold: it increases the cross section and it lowers the threshold from $2m$ the monopole-antimonopole production threshold. To observe the other dependence, in Fig.~\ref{Mmonopolemasses} we fix the binding energy to $100$ GeV and vary the monopole mass. The effect goes inversely proportional to the monopole mass, i.e., the lower the monopole mass the larger the  cross sections. 

%%%%%%%%%%%%%%%%%%%%%%%%%%%%%%%%%%%%%%%%%%%%%%%%%%%%%%%%
%          Fig. 20  variation with fixed monopole mass and varying Ebind
%%%%%%%%%%%%%%%%%%%%%%%%%%%%%%%%%%%%%%%%%%%%%%%%%%%%%%%%
%%%%%%%%%%%%%%%%%%%%%%%%%%%%%%%%%%%%%%%%%%%%%%%%%%%%%%%%

%%%%%%%%%%%%%%%%%%%%%%%%%%%%%%%%%%%%%%%%%%%%%%%%%%%%%%%%
%          Fig. 21  different monopole masses and fixed Ebind
%%%%%%%%%%%%%%%%%%%%%%%%%%%%%%%%%%%%%%%%%%%%%%%%%%%%%%%%
%%%%%%%%%%%%%%%%%%%%%%%%%%%%%%%%%%%%%%%%%%%%%%%%%%%%%%%%

\begin{figure}[htb]
\begin{center}   
\includegraphics[scale=1.2]{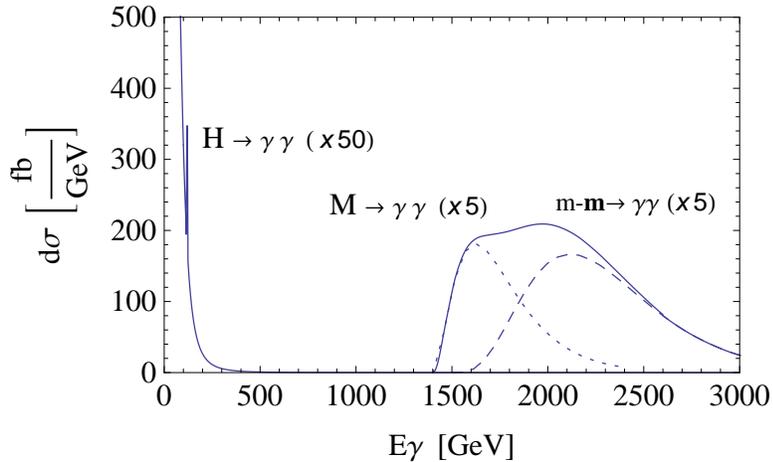}
\caption{We show the differential two photon crossection with the Higgs signal (scaled by a factor of 50 over the background) where the monopolium and monopole-antimonopole contributions have been incorporated. We also show the monopolium signal (dotted) that presents a sudden rise at threshold, and the monopole- antimonopole  (dashed)  with a softer increase at threshold. Note that both signals were multiplied by 5 . The mass of the monopole in both cases is $750$ GeV and that of the monopolium $1400$ GeV. The background has been obtained from ref. \cite{atlas1}. }
\label{mmMhiggs}
\end{center}
\end{figure}

We can summarize our findings by stating  that low monopole masses and large bindings favor the detection of monopolium. Monopolium has the advantage over the monopole-antimonopole process of lowering the threshold, narrowing the bump and  increasing the cross section with binding energy. In Fig.~\ref{mmMhiggs} we show three interesting effects, namely the two photon decays of Higgs,  monopolium and monopole-antimonopole annihilation. The Higgs signal has been increased over the background by a factor of 50. The parameters for the monopolium cross section shown are $m=750$ GeV and $ M= 1400$ GeV, which have been chosen so that its signal is of the same size of that of the monopole-antimonopole annihilation  with a monopole mass of also $750$ GeV.  We note  two  of the features mentioned before, the lower threshold and the narrower bump structure. If we would increase the binding by a few tens of GeV the height of the bump would increase considerably with respect to the monopole-antimonopole cross section (recall Fig.~\ref{MggEbind}). Note that the $\gamma \gamma$ conventional background at the monopole scenario in this kinematical region is extremely small, as measured recently by ATLAS in diphoton studies
\cite{diphoton} and while searching for $H\rightarrow\gamma\gamma$~\cite{hgg}.

%%%%%%%%%%%%%%%%%%%%%%%%%%%%%%%%%%%%%%%%%%%%%%%%%%%%%%%%
%          Fig. 22  Higgs-mm-M 
%%%%%%%%%%%%%%%%%%%%%%%%%%%%%%%%%%%%%%%%%%%%%%%%%%%%%%%%
%%%%%%%%%%%%%%%%%%%%%%%%%%%%%%%%%%%%%%%%%%%%%%%%%%%%%%%%

If  a broad bump would appear experimentally it could certainly arise from monopole dynamics. Thereafter  the way to distinguish between the decay of monopolium and the annihilation of monopole-antimonopole would be through the angular
dependence. In the present scenario, a spin zero monopolium, the angular decay properties would be similar to that of para-positronium \cite{Adkins:2001zz}, while that of monopole-antimonopole would be analogous to light-by-light scattering in QED \cite{Karplus:1950zz,Csonka:1974ey}. Moreover, the two pehenomena could occur simultaneausly, as happens in the case of electrons and positrons, where we have in light-by-light scattering electron-positron annihilation and positronium decay. If the latter were the case two bumps, if the overlap is not large, or a very broad flat bump, if the overlap is considerable, could be seen. The existence of one or two bumps depends very strongly on the binding dynamics and the monopole mass. 
Note that there is no possible confusion with the Higgs, since its width is narrow compared to its mass. Moreover, the scattering cross section
for a heavy Higgs in the two photon channel is extremely small compared to that in the other channels, and therefore, its characteristics would be known by the time a two photon bump
would be seen.

\section{Conclusions}\label{conclusions}

The existence of Dirac monopoles would modify our understanding of QED. The DQC is a beautiful consequence of the existence of monopoles and therefore they represent an extremely appealing physical scenario. There is as of yet no experimental proof of their existence. This has led to approximate mass bounds which suggest a mass scale for the monopole above 500 GeV. LHC opens up this energy regime for research and therefore monopoles become again a subject of exciting experimental search.

The DQC, implying a huge magnetic coupling constant, complicates matters from the theoretical point of view. Non perturbative methods are required. We have avoided the problem by using an effective theory valid at one loop order. In our scheme, the effective coupling $\beta g$  is still large away from threshold, but the expected high monopole masses provide a convenient cut-off which makes the theory meaningful.  In the context of this theoretical scheme we are able to calculate monopole(antimonopole) production and monopole-antimonopole annihilation. Centering our description to LHC, a proton machine,  we have described scenarios for the production of monopole-antimonopole pairs via photon fusion. We have described their annihilation into photons via the conventional  box diagram, analogous to that of light-by-light scattering in conventional QED.  The only difference in our approach is the introduction of a threshold factor in the form of a velocity. We have analyzed in detail the elementary process, light-by-light scattering at monopole energies, and have subsequently incorporated the description of the photon-photon elementary scattering into the initial proton-proton collision. One main assumption is that no other particle contributes to the box diagram in the region of interest. We have chosen in the present calculation a mass range which is above conventional particle masses and below the supersymmetric particle ranges. One could think of interference with  elementary particle decays, however the latter would have a resonant structure which our box diagram does not have. Thus any interference would be avoided by their different geometric structure.

With all these preventions we have shown that monopoles up to 1 TeV in mass should be detected at LHC in the $\gamma \; \gamma$ channel at present running energies and with present luminosities. The signature is a  rise of the cross section which should stay for a long energy range, because no resonant peak structure is to be expected.

Even if monopoles exist it might be possible that due to the very strong magnetic coupling they do not appear as free states but bound forming 
monopolium, a neutral state,  very difficult to detect directly. We have  analyzed the coupling of monopolium to photons and its contribution 
to light-by-light like scattering, i.e. monopolium's dynamical decay. We have found that for reasonable values of the monopole mass and relatively 
small, compared with their mass, binding energies, spin-zero monopolium disintegrates into two gammas with cross sections which are reachable with the currently available LHC integrated luminosity of 5 fb$^{-1}$.

Our investigations go beyond this wishful scenario. We have seen that the cross section depends both on the mass on the monopole and on that 
of monopolium, increasing inversely proportional to the monopole mass and being directly proportional to the binding energy. This means that 
similar cross section can be achieved with very heavy monopoles if the binding energy is large. 

To conclude, without  doubt, the appearance of a broad signal at threshold,  in the two photon cross section, should be considered as a clear manifestation of the monopole presence. The implementation of our findings in detector analysis would provide actual observations.

\section*{Acknowledgement}
We thank the authors of JaxoDraw  for making drawing diagrams an
easy task \cite{Binosi:2003yf}.  LNE, HF
and CAGC were partially supported by CONICET and ANPCyT Argentina. VAM acknowledges support by the Spanish Ministry of Science and Innovation (MICINN) under the project FPA2009-13234-C04-01, by the Spanish Agency of International Cooperation for Development under the PCI projects A/023372/09 and A/030322/10 and by the CERN Corresponding Associate Programme. VV has been supported  by HadronPhysics2,  by  MICINN (Spain) grants FPA2008-5004-E,  FPA2010-21750-C02-01,   AIC10-D-000598 and by GVPrometeo2009/129.

\end{document}